\documentstyle[aasms4]{article}
\newcommand{\beq}{\begin{equation}}
\newcommand{\eeq}{\end{equation}}
\newcommand{\etal}{et~al.}
\newcommand{\kms}{km s$^{-1}$}

\begin{document}

\title{
Interferometric Astrometry of \\ the Detached White Dwarf - M Dwarf Binary Feige 24\\ Using {\it 
Hubble Space Telescope} 
Fine
Guidance Sensor 3: \\
White Dwarf Radius and Component Mass Estimates \footnote{Based on 
observations made with
the NASA/ESA Hubble Space Telescope, obtained at the Space Telescope
Science Institute, which is operated by the
Association of Universities for Research in Astronomy, Inc., under NASA
contract NAS5-26555} }

\author{ G.\ Fritz Benedict\altaffilmark{1}, Barbara E.
McArthur\altaffilmark{1},
Otto G. Franz\altaffilmark{2}, L.\ 
H. Wasserman\altaffilmark{2}, E.\ Nelan\altaffilmark{5}, J. 
Lee\altaffilmark{7}, \\ L.W.\
Fredrick\altaffilmark{12}, W.\ H.\ Jefferys\altaffilmark{6}, 
W.~van~Altena\altaffilmark{7}, E. L. Robinson\altaffilmark{6}, W. J. Spiesman\altaffilmark{1}, \\P.~J.~Shelus\altaffilmark{1},
 P.D. Hemenway\altaffilmark{8}, R. L.
Duncombe\altaffilmark{9}, D. Story\altaffilmark{10}, A.\ L.\
Whipple\altaffilmark{11}, and A. Bradley\altaffilmark{11}}

\altaffiltext{1}{McDonald Observatory, University of Texas, Austin, TX 78712}
\altaffiltext{2}{Lowell Observatory, 1400 West Mars hill Rd., Flagstaff, AZ 
86001}
\altaffiltext{5}{Space Telescope Science Institute, 3700 San Martin Dr., 
Baltimore, MD 21218}
\altaffiltext{6}{Astronomy Dept., University of Texas, Austin, TX 78712}
\altaffiltext{7}{Astronomy Dept., Yale University, PO Box 208101, New Haven, CT 
06520}
 \altaffiltext{8}{Oceanography, University of Rhode Island, Kingston, RI 02881}
\altaffiltext{9}{Aerospace Engineering, University of Texas, Austin, TX 78712}
\altaffiltext{10}{ Jackson and Tull, Aerospace Engineering Division
7375 Executive Place, Suite 200, Seabrook, Md.  20706}
\altaffiltext{11}{Allied-Signal Aerospace, PO Box 91, Annapolis Junction, MD 
20701}
\altaffiltext{12}{Astronomy Dept., University of Virginia, PO Box 3818, 
Charlottesville, VA 22903}



\begin{abstract}
With {\it HST} FGS 3 we have determined a parallax for the white dwarf - M dwarf 
interacting binary, Feige 24. The white dwarf (DA) component has an effective 
temperature, $T_{eff} \sim 56,000 K$. A weighted average with past parallax determinations 
($\pi_{abs} = 14.6 \pm 0.4$ milliseconds of arc) narrows the range of possible radius 
values, compared to past estimates. We obtain $R_{DA} = 0.0185 \pm 0.0008 R_{\sun}$ with uncertainty in the temperature and bolometric correction the dominant contributors to the error. FGS photometry provides a light curve entirely consistent with reflection effects. A recently refined model Mass-Luminosity Relation (Baraffe et al. 1998) for low mass stars provides a mass estimate for the M dwarf companion, ${\cal M}_{dM} = 0.37 \pm 0.20 {\cal M}_{\sun}$, where the mass range is due to  metallicity and age uncertainties. Radial velocities from Vennes and Thorstensen (1994) provide a mass ratio from which we obtain ${\cal M}_{DA} = 0.49^{+0.19}_{-0.05} ~{\cal M}_{\sun}$. Independently, our radius and recent $log~g$ determinations yield $0.44 < {\cal M}_{DA} < 0.47 {\cal M}_{\sun}$. In each case the minimum DA mass is that derived by Vennes \& Thorstensen from their radial velocities and Keplerian circular orbits with $i \le 90 \deg$. Locating Feige 24 on an ${\cal M} - R$ plane suggests a carbon core. 

Our radius and these mass estimates yield a $\gamma_{grav}$ inconsistent with
that derived by Vennes \& Thorstensen. We speculate on the nature of a third component whose existence would resolve the discrepancy.
\end{abstract}




%

\section{Introduction}

Feige 24 ( = PG 0232+035 = HIP 12031) is DA white dwarf, red dwarf (M1-2V) (\cite{Lie77}) binary (P = 4.23 days, \cite{Ven94}=VT94) that is 
described as 
the prototypical post-common envelope detached system with a low probability of becoming a Cataclysmic Variable (CV) within a Hubble time (\cite{Kng94} and Marks, 1994).
This object was selected for our {\it HST} parallax program because a directly 
measured distance 
could reduce the uncertainty of the radius of one of the hottest white dwarfs. 
Since the instigation of this 
program and the selection of targets over 15 years ago, at least two other 
groups have measured a parallax for Feige 24; USNO-Flagstaff (\cite{Dah88}), and 
{\it 
HIPPARCOS} (\cite{Per97} and \cite{Vau97}). We outlined the results of a preliminary analysis
in \cite{Ben00}. Here we discuss our analysis and final results in detail.

\cite{Pro98} presented radii derived from {\it HIPPARCOS} parallaxes for 21 
white dwarfs. In most cases, the dominating error term for the white dwarf radii 
was the parallax uncertainty. Our parallax of Feige 24, while slow in coming, 
has provided a fractional parallax uncertainty,
$\Delta\pi\over\pi$, similar to those in the  \cite{Pro98} study, but for a much 
hotter, more distant object.

We time-tag our data with 
a modified Julian Date, $MJD 
= JD - 2400000.5$. We abbreviate millisecond of arc, mas; white dwarf, DA; and M 
dwarf, dM, throughout.

\section{The Astrometry}
Our astrometric observations  were
obtained with Fine Guidance Sensor 3 (FGS 3), a two-axis, white-light
interferometer  aboard {\it HST}. 
\cite{Bra91}
provide an overview of the
FGS 3 instrument and \cite{Ben99} describe the astrometric capabilities 
of FGS 3 and typical data acquisition and reduction strategies.

\begin{figure}
\epsscale{1.0}
\plotone{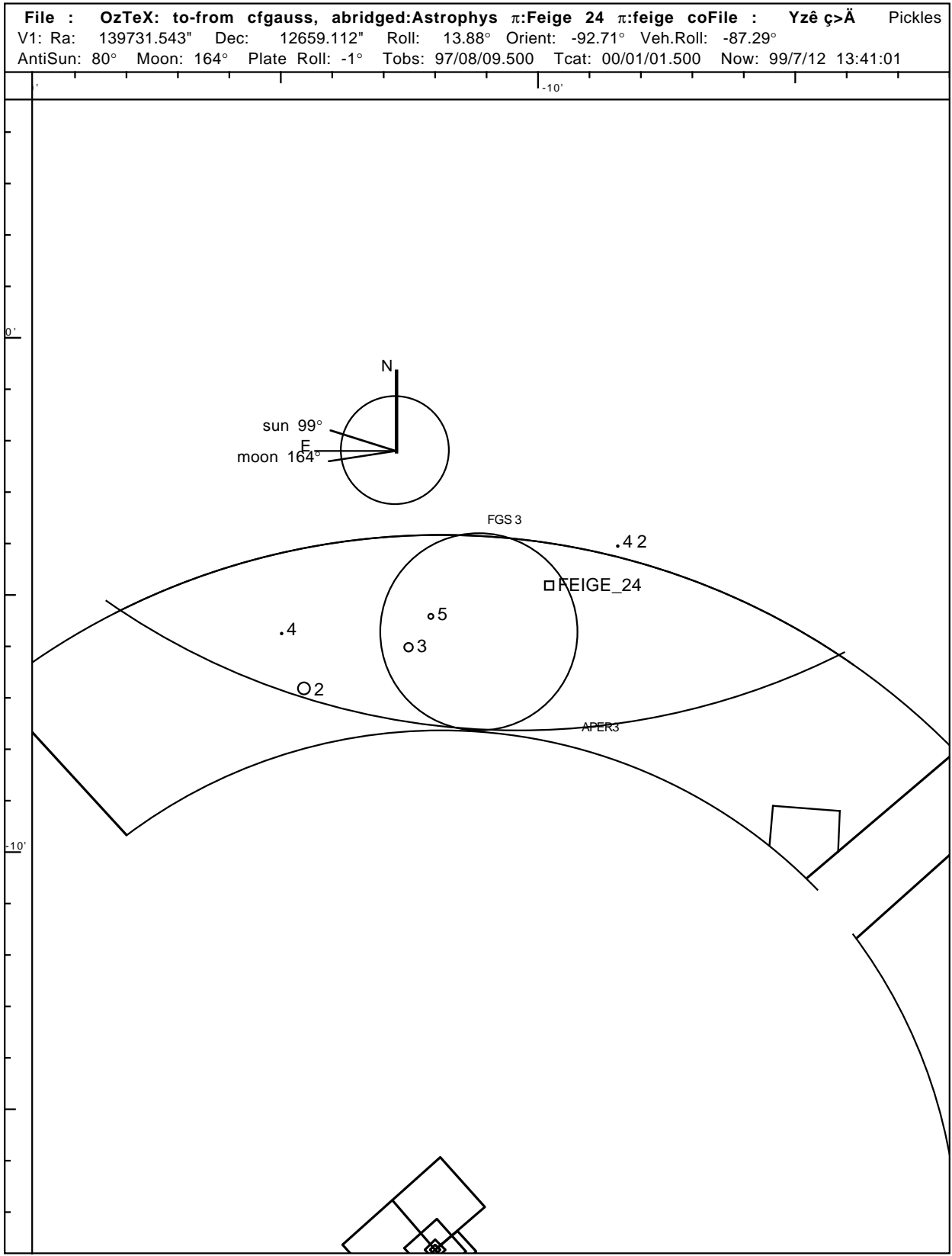}
\caption{Location of reference stars within the FGS 3 field of regard
on 08 Aug 1997. Note the less than ideal placement of the primary science target 
with respect to the reference frame.} 
\label{fig-1}
\end{figure}

We use the term `pickle' to describe the field of regard of the FGS.
The instantaneous field of view of FGS 3 is a
$5 \times 5$ arcsec square aperture.  Figure~\ref{fig-1} shows a finding
chart for Feige 24 and our astrometric reference stars in the  FGS 3 pickle as
observed on 08 Aug 1997. Note the less than ideal placement of the primary science target 
with respect to the reference frame. The placement of Feige 24, at one side of the distribution of reference stars, seems to have produced no adverse astrometric or photometric effects.

\subsection{The Astrometric Reference Frame}  \label{AstRefs}
Table~\ref{tbl-4} provides a list of the observation epochs. Our data reduction 
and calibration procedures are described in \cite{Ben99} and \cite{Mca99}.
We obtained a total of 71 successful measurements of our reference 
stars during eight 'observing runs'. For each of these eight observation sets we determine the scale and rotation relative to the sky, using a 
GaussFit (\cite{Jef88}) model. The orientation of the observation sets is obtained 
from ground-based astrometry (USNO-A2.0, \cite{Mon98}) with uncertainties in the 
field orientation $\pm 0\fdg12$.  

Having only 8 observation sets and four 
reference stars precludes us from our 
usual practice (\cite{Ben99}) of constraining the proper motions and parallaxes 
to sum to zero ($\Sigma \mu = 0$ and $\Sigma \pi = 0$) for the entire reference 
frame. 
From a series of solutions we determined that only reference star ref-3 has a 
statistically significant proper motion and parallax. So, we constrain $ \mu = 
0$ and   
$ \pi = 0$ for reference stars ref-2, -4 and -5.

\begin{figure}
\epsscale{0.5}
\plotone{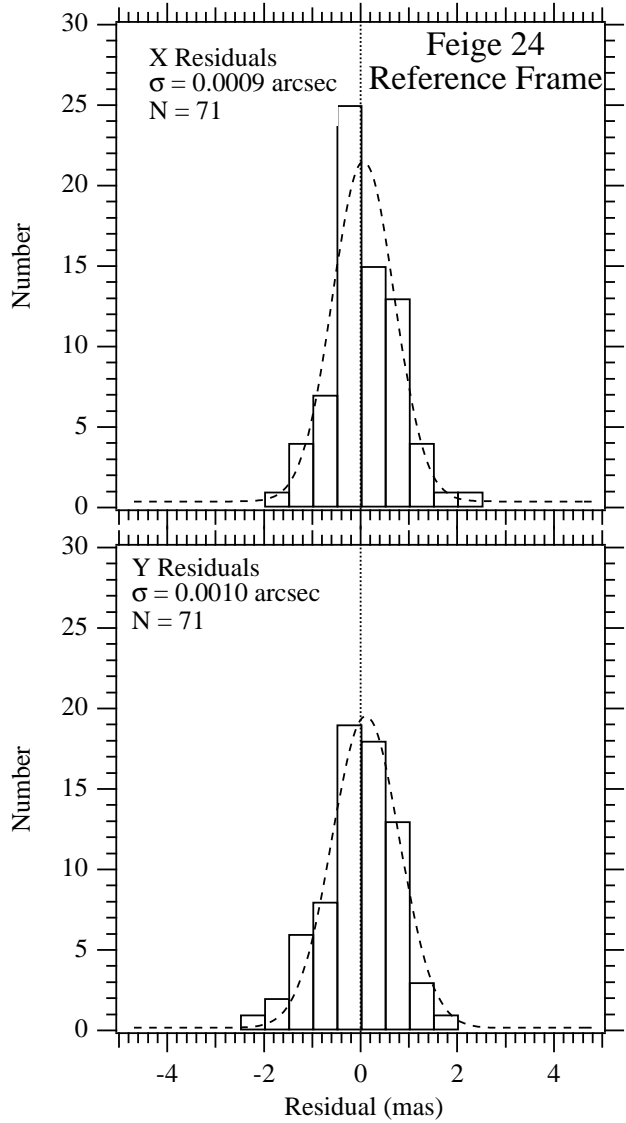}
\caption{Histograms of x and y residuals obtained from modeling the Feige 24 
reference frame to obtain scale, orientation, and offset parameters. 
Distributions are fit with gaussians.} 
\label{fig-2}
\end{figure}

We conclude from histograms (Figure~\ref{fig-2}) of the reference star residuals 
that we have obtained 
a per-observation precision of $\sim 1$ mas. The resulting reference frame 
'catalog' (Table~\ref{tbl-3}) 
was determined
with final errors	$<\sigma_\xi > = 0.5$	 and	$<\sigma_\eta >= 0.6$ mas. 

To determine if there might be unmodeled, but eventually correctable, systematic 
effects at the 1 mas level, we plotted the Feige 24 reference frame X and Y 
residuals against a number of spacecraft, instrumental, and astronomical 
parameters. These included X, Y position within the pickle; radial distance from 
the pickle center; reference star V magnitude and B-V color; and epoch of 
observation.  We saw no trends, other than the expected increase in 
positional uncertainty with reference star magnitude. 
\setcounter{footnote}{0}
\subsection{Modeling the Parallax and Proper Motion of Feige 24}
Spectroscopy of the reference frame stars obtained from the WIYN \footnote{The 
WIYN Observatory is a joint facility of the University of
Wisconsin-Madison, Indiana University, Yale University, and the National
Optical Astronomy Observatories.} and an estimate of color excess, E(B-V), from
\cite{Bur82}. (Table~\ref{tbl-3}) shows that the colors of 
the reference stars and our science target differ, with $\Delta(\bv) \sim -1$. 
Therefore,  we apply the differential correction for lateral color discussed in 
\cite{Ben99} to the Feige 24 observations and obtain
a parallax relative to our reference frame, $\pi_{rel} = 13.8 \pm 0.4$ mas. The proper motion relative to the four astrometric reference stars is listed in Table~\ref{tbl-5}.

\begin{figure}
\epsscale{0.5}
\plotone{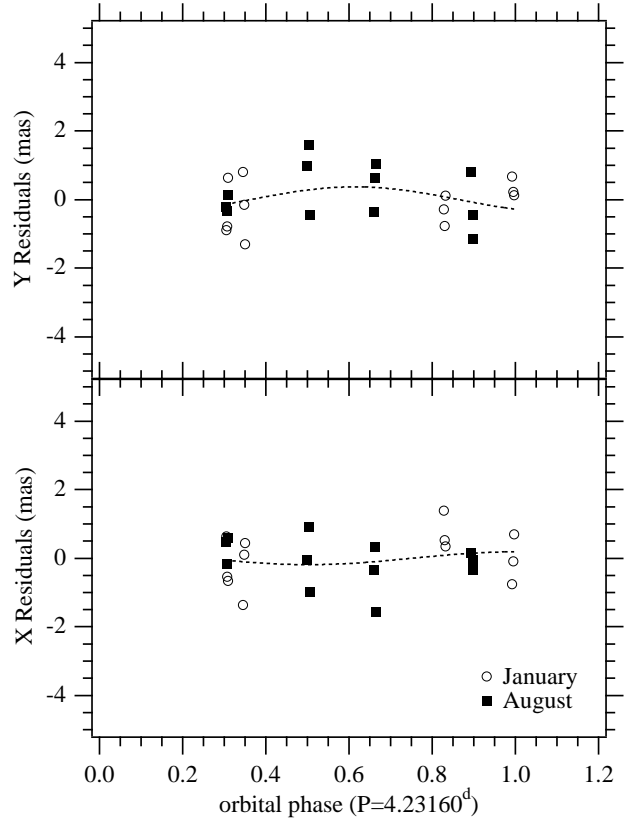}
\caption{Astrometric residuals in RA (X) and Dec (Y) phased to the VT94 orbit ($P = 4.23160^d$ and
$T_0 = JD 2448578.3973$). Boxes and open circles denote the two {\it HST} orientations, which seem to have no effect on the astrometric residuals. Dashed lines are best-fit sine waves constrained to the VT94 period.} \label{fig-res}
\end{figure}

\cite{Fra98} and \cite{Ben99} have demonstrated 1 mas astrometric precision for FGS 3. Table \ref{tbl-4} presents our Feige 24 astrometric residuals obtained from the parallax and proper motion model. Histograms of these residuals are characterized by $\sigma_x = 1.0$ and $\sigma_y = 1.2$ mas. This was slightly larger than expected. To investigate whether or not the larger residuals could be attributed to Feige 24, Figure~\ref{fig-res} presents the residuals phased to the VT94 orbital period, $P = 4.23160^d$, with $T_0 = HJD~ 2448578.3973$. We find no significant trends in the astrometric residuals. In particular, there is no correlation with the two distinct {\it HST} orientations required by the pointing constraints discussed in \cite{Ben99}. With any reasonable masses for the DA and dM components, a binary system at this distance, having this period, could exhibit maximum reflex motion at the 0.5 mas level. This null detection does not place very useful upper limits on the component masses.

Because our parallax for Feige 24 is
determined with respect to the reference frame stars which have their own
parallaxes, we must apply a correction from relative to absolute parallax.
The WIYN spectroscopy and the estimated color excess (See Table 2)
indicate a reference frame with an average parallax of  $<\pi >_{ref} =0.9 \pm 0.4$ mas. where the error is based on the dispersion of the individual
spectrophotometric parallaxes.  To check our correction to absolute, we
compare it to those used in the Yale Parallax Catalog (YPC95, van Altena, Hoffleit, \& Lee, Section 3.2). 
From YPC95, Fig. 2, the Feige 24 galactic latitude, $b = -50\fdg3$ and 
average magnitude for the reference frame, $<V_{ref}>=13.4$, we obtain a 
correction to absolute of 1.9 mas. Rather than use a galactic model-dependent 
correction, we adopt the spectroscopically derived 
$<\pi >_{ref} =0.9 \pm 0.4$ mas. Applying this correction results in an
absolute parallax of $\pi_{abs} = +14.7 \pm 0.6$ mas, where the error has equal
contributions from the HST FGS observations and the correction to absolute
parallax.  Finally, we note that our proper motion is smaller than either
the HIPPARCOS or USNO values, where the HIPPARCOS value is an absolute
proper motion while the USNO and the HST values are relative to their
respective reference frame proper motions.  If our reference stars are a
representative statistical sample of the parent population, then based on
the data in Table III in van Altena (1974), we expect a
statistical uncertainty in the mean value of the correction to absolute 
proper motion (not applied here) of $\pm 6$ mas y$^{-1}$.

\begin{figure}
\epsscale{1.0}
\plotone{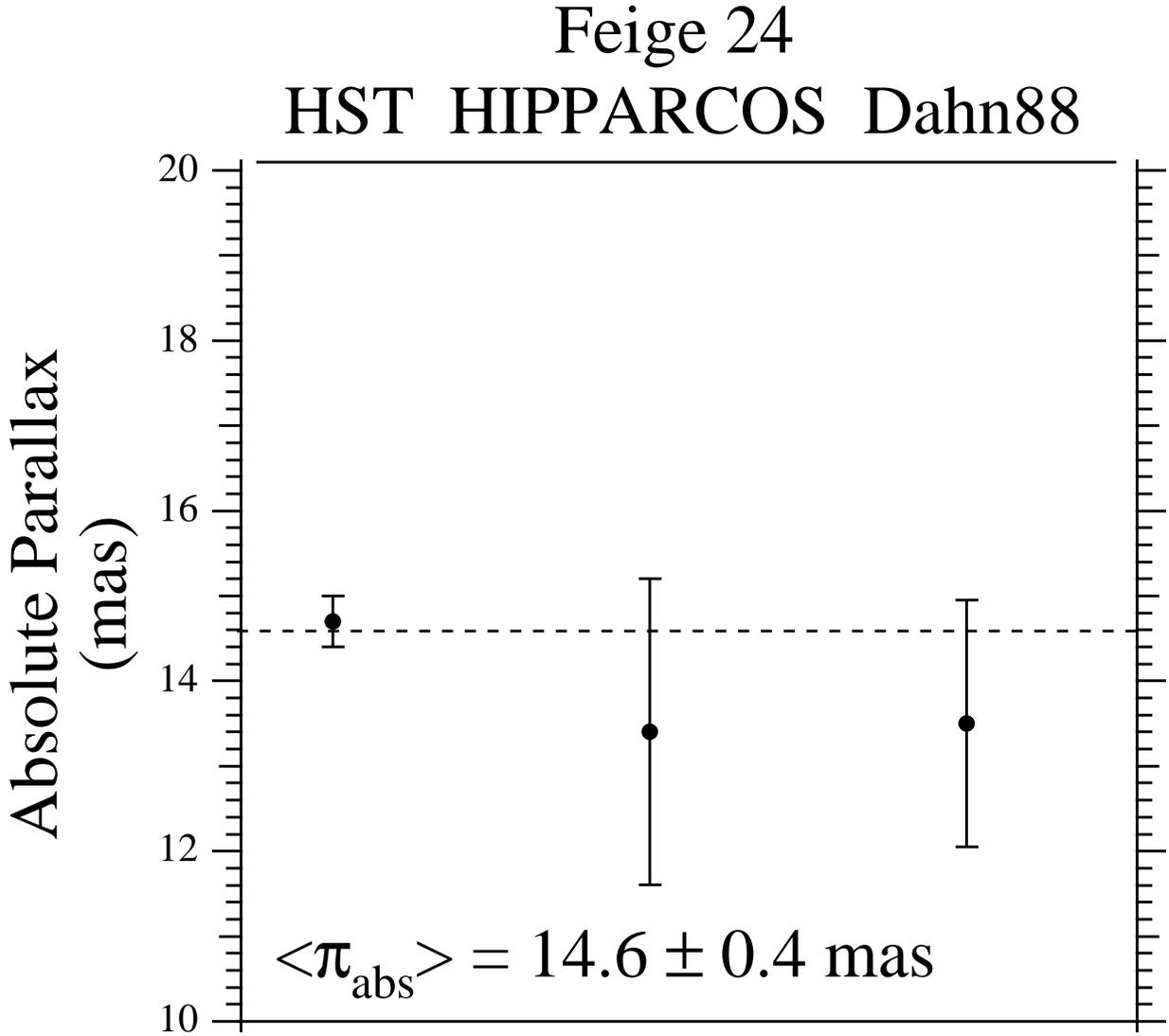}
\caption{Absolute parallax determinations for Feige 24. We compare
{\it HST}, {\it HIPPARCOS}, and USNO (\cite{Dah88}). Error bars are 1-$\sigma$. The horizontal dashed line 
gives the weighted average absolute parallax, $<\pi_{abs}>$.} 
\label{fig-3}
\end{figure}

We compare our absolute parallax to previous work in Table~\ref{tbl-5} and in 
Figure~\ref{fig-3}. We adopt for the remainder of this paper the weighted 
average absolute parallax, $<\pi_{abs}> = 14.6 \pm 0.4$ mas,
shown as a horizontal dashed line in Figure~\ref{fig-3}. Weights used are 
$1/\sigma^2$.

Lutz \& Kelker (1973) show that for a uniform distribution of stars, the
measured trigonometric parallaxes are strongly biased towards the observer 
(i.e., too large), rendering inferred distances and luminosities too small. This 
bias is proportional to $(\sigma_\pi/\pi)^2$. Using a space 
density determined for the CV RW Tri (McArthur et al.
1999),  and presuming that Feige 24 is a member of that
same class of object (binaries containing white dwarfs), we determine an LK correction of $-0.01 \pm 0.01$ 
magnitudes.  Correcting our distance modulus, we obtain $m-M = 4.17 \pm 0.11$.

\subsection{Kinematic Age of the Feige 24 System}
From the VT94 systemic radial velocity and either our proper motions or those from HIPPARCOS (Table \ref{tbl-5}) we derive the space velocity of Feige 24,  67 $\pm$ 1 \kms. 
The velocity component perpendicular to the galactic plane, W,  is -37 \kms.  Our new parallax places the star 53 parsecs
below the Sun or 61 parsecs below the galactic 
plane.  An object this far below the galactic plane and continuing to move further away from the plane so swiftly is more characteristic of a 'thick disk' than a thin disk object (c.f. \cite{The97}). Feige 24, if truly a Pop I object, has a space velocity 3.5 times the young disk velocity dispersion. These data suggest that Feige 24 formed prior to the 
formation of the galactic disk, although subsequent evolution of the 
DA component is likely quite recent.  This may be an instance of
past mass transfer in an intermediate Population II object. 

\section{Astrophysics of the Feige 24 System}
We discuss the consequences of a more precisely determined parallax, calculating 
some astrophysically relevant parameters for the DA and dM components. These are 
collected in Table~\ref{tbl-6}. Our goals are the radius and mass of the DA 
component. We first calculate a radius, then estimate the time since the DA 
formation event.
Component masses have been estimated by VT94. We will revisit this issue later. That we do 
not substantially improve the mass uncertainty motivates a future direct 
measurement of the component separation. This one measurement would yield 
precise masses. A series of measurements would provide individual orbits, 
possibly illuminating past and future component interactions.
\subsection{Estimating the DA Radius}
To estimate the DA radius we require an intrinsic luminosity. From \cite{Lan83} 
we obtain a system total magnitude, $ V_{tot }= 12.41 \pm 0.01 $. 
The magnitude of the white dwarf is critical and difficult to obtain, because the M dwarf always contributes flux. \cite{Hol86} derive $V_{DA} = 12.56 \pm 0.05$ using IUE spectra. They ratio 
Feige 24 with other hot DA; G191 B2B, GD246 and HZ43. From the DA magnitude and total magnitude we 
obtain $V_{dM} = 14.63 \pm 0.05$ and $\Delta V = 2.07$. We assume an $A_V = 0$ 
for Feige 24 at d = 69 pc, consistent with our adopted $A_V = 0.09$ for the 
reference frame at an average distance d = 1600 pc (Table~\ref{tbl-3}). The LK 
bias-corrected distance modulus ($m-M = 4.17 \pm 0.11$) then yields absolute magnitudes $M_V = 10.46 \pm 
0.12$ for the red dwarf companion and $M_V = 8.39 \pm 0.12$ for the DA.
 
A recently 
determined temperature of the Feige 24 DA, taking into account non-LTE and heavy element effects (\cite{Bar98}), is $T_{eff}^{DA} = 56,370 \pm 1,000 K$. This temperature yields a radius via differential comparison with the 
sun. This procedure requires a bolometric magnitude, hence, a bolometric correction. We could adopt the bolometric correction, B.C. = -4.88, generated by Bergeron et al. (1995) from 
a pure Hydrogen, $log~g = 8$ DA model convolved with a V bandpass. But, Feige 24 is neither $log~g = 8$ nor pure H. 

Flower (1996) provides bolometric corrections for normal stars up to $T_{eff} \sim 54,000K$. From Flower (1996), figure 4, the relationship between $log T_{eff}$ and B.C. is linear for  $T_{eff} > 25,000 K$. Hotter stars lie on the Rayleigh-Jeans tail of the blackbody curve, where flux is roughly proportional to $T_{eff}$, not $T_{eff}^4$. A small linear extrapolation yields B.C. = -4.82$\pm 0.06$ for the Feige 24 DA. The B.C. error comes from the uncertainty in $T_{eff}^{DA}$. 
Because a DA with some heavy elements in its atmosphere radiates more like a hot normal star than a pure H DA, we choose the Flower correction rather than the model correction. We are also encouraged by the near equality of the B.C. from observation and theory.

We obtain a DA 
bolometric luminosity $M_{bol}^{DA}  = M_V + B.C. = 3.57 \pm 0.13$. $R_{DA}$ follows from the expression
\beq
M_{bol}^{\sun} - M_{bol}^{DA} = 10~log(T_{eff}^{DA}/~T_{eff}^{\sun})+ 5~log(R_{DA}/R_{\sun})
\eeq
where we assume for the Sun $M_{bol}^{\sun}=+4.75$ and $T_{eff}^{\sun}=5800 K$. 
We find $R_{DA} = 0.0180 \pm 0.0013 R_{\sun}$, following the error analysis of 
\cite{Pro98}. The primary sources of error for this radius are the 
bolometric correction and the $T_{eff}^{DA}$. 

A second approach to deriving $R_{DA}$ involves the V-band average flux, $H_V$, discussed in 
Bergeron et al. (1995). They list $H_V^{DA}$ as a function of temperature for, again, the pure Hydrogen, $log~g = 8$ model. If we can determine an $H_V^{\sun}$, we can derive $R_{DA}$ from
\beq
R_{DA} = (H_V^{\sun}/H_V^{DA})10^{-0.4(M_V^{DA}-M_V^{\sun}})
\eeq
where $M_V^{DA} = 8.39 \pm 0.12$ comes from our parallax and $M_V^{\sun} = 4.82$ is assumed. We obtain $H_V^{\sun}$ by convolving the Bessel (1990) V band response with the solar spectral distribution listed in Allen (1973). We calculate $H_V^{\sun} = 6.771\times10^5$ ergs cm$^{-2}$ s$^{-1}$ \AA$^{-1}$ str$^{-1}$.
We obtain for $T_{eff}=56,370 K$ an $R_{DA}=0.0188 \pm 0.0010$. 

A weighted average of the two independent determinations provides $R_{DA} = 0.0185 \pm 0.0008 R_{\sun}$, where the error is certainly underestimated due to unknown systematic effects. Parallax is no longer a significant source of error for the radius determination.
Comparing with the results presented in \cite{Pro98}, figure 7, we find Feige 24 
to have a radius larger than any other white dwarf.

With a temperature $T_{eff} \sim 56,000$, the time since 
the DA formation event is unlikely to be longer than 1.5 My. This conclusion is 
drawn from the  DA cooling tracks as function of mass calculated by M. Wood, 
detailed in \cite{Sio99}, fig. 7. These models also indicate that the DA mass 
must satisfy ${\cal M}_{DA} \ge 0.4 {\cal M}_{\sun}$ to remain near this lofty 
$T_{eff}$ for longer than $3\times10^5$ y.

\subsection{Estimating The White Dwarf Mass}
Before estimating ${\cal M}_{DA}$ we review the VT94 minimum component masses from their radial velocities and the Kepler relation for total system mass, separation, and period. Then, we estimate the DA mass using two different approaches. We first attempt to determine the most likely dM mass. The VT94 radial velocity amplitude ratio then provides the DA mass. The second, independent mass estimate follows from our derived radius along with the DA atmospheric parameter, $log~g$, obtained through spectroscopy. Our DA mass estimate will differ little from VT94, and, if better, is so only by virtue of more recent dM models and DA atmospheric parameters.

\subsubsection{Minimum Component Masses from Binary Radial Velocities}
The system total lower mass limit can be set by the VT94 radial velocities and the Kepler relation for mass, separation, and period. VT 94 give us the velocities along each component orbit, the fact that each orbit is circular (from the pure sine wave fits to the velocity curves), and the period, the time it takes to travel around each orbit. Assuming an edge-on system ($i = 90 \deg$), one that can produce the full vector amount of radial velocity amplitude measured by VT94, the minimum system mass is ${\cal M}_{tot} = 0.73 {\cal M}_{\sun}$. From the VT94 mass ratio, ${\cal M}_{dM}/{\cal M}_{DA}=0.63 \pm 0.04$, we obtain the DA mass limit, ${\cal M}_{DA~Kepler} \ge 0.44{\cal M}_{\sun}$, and the dM mass limit, ${\cal M}_{dM~Kepler} \ge 0.26{\cal M}_{\sun}$. No smaller masses can produce the observed radial velocities for orbits of these known sizes. 
At d = 68.5 pc an edge-on system with minimum mass would separate the components by 672 microarcsec, or $9.9R_{\sun}$.

\subsubsection{Inclination from the Light Curve}

\begin{figure}
\epsscale{0.5}
\plotone{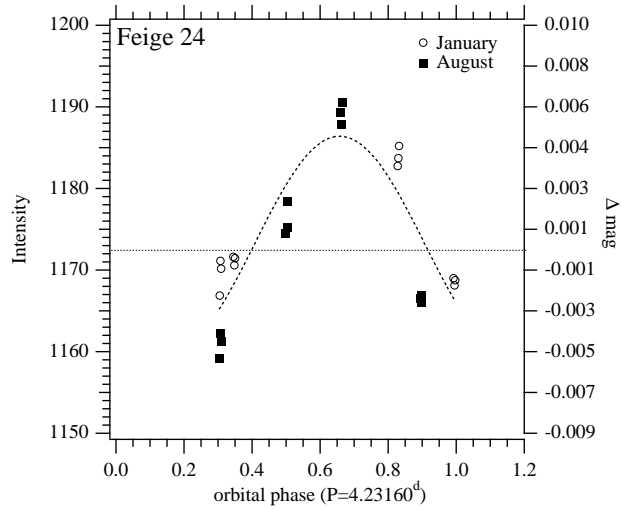}
\caption{Flat fielded intensity  (the filter (F583W) has a bandpass
centered on 583 nm, with 234 nm FWHM) and differential instrumental magnitudes
phased to the VT94 orbit ($P = 4.23160^d$ and
$T_0 = JD 2448578.3973$). Boxes and open circles denote the two {\it HST} orientations, which seem to have no effect on the photometry. The dashed line is a best-fit sine wave constrained to the VT94 period.} \label{fig-LC}
\end{figure}

VT 94 find H$\alpha$ equivalent width variations that phase with the orbital period. These show a maximum at $\phi = 0.5$. Photometric variations of Feige 24 might be detectable, because the photometric capabilities of FGS 3 approach a precision of 0.002 magnitude (\cite{Ben98}). Figure~\ref{fig-LC} shows the flat-fielded counts and the corresponding differential instrumental magnitudes as well as a sin
wave fit with amplitude and phase as free parameters. There is a clear photometric signature with a peak-to-peak amplitude 0.028 magnitude, showing maximum system brightness at phase $\phi = 0.58 \pm 0.09$. Given the sparse coverage, this phase at maximum is not surprisingly different from the H$\alpha$ equivalent width maximum seen at $\phi = 0.5$. 

A likely mechanism for producing the single-peaked orbital light 
curve is heating 
of the dM star by the white dwarf (the reflection effect).
As the dM star orbits the white dwarf, its heated face is alternately
more or less visible, increasing and decreasing the observed flux
from Feige 24 once per orbit.
To test this hypothesis we calculated model light curves 
using an updated version of the light curve synthesis program
described by Zhang et al.\ (1986).
We initially adopted $T_{eff} = 56,370$~K and $R = 0.0185\ R_\odot$ for the
white dwarf, $T_{eff} = 3800$~K and $R = 0.52\ R_\odot$ for the
M1-2V star, and $4.8 \times 10^{-2}$~AU ($10.3R_{\sun}$) for the separation of their
centers of mass, and then adjusted the temperature of the dM star
so that it contributed 13.5\% of the V flux from the system.
The peak-to-peak amplitudes of the resulting model light curves are a 
function of orbital inclination, topping out at $\sim0.025$ mag for
$i = 90^\circ$, and can easily be made to agree in 
amplitude and shape with the observed light curve.

This photometric behavior is entirely consistent with reflection effects (c.f. Robinson et al. 2000).
We find that the quality of the observed light curve is, however, 
inadequate to improve the parameters of the system, particularly the inclination. We have not sufficiently sampled the expected flat section of the light curve (near $\phi = 0$).
Nevertheless, these results do provide quantitative evidence that
(1) the orbital light curve is caused by heating and
(2) the heating is consistent with the radius and temperature we
have derived for the white dwarf -- a useful external check on our results.

\subsubsection{DA Mass from the M Dwarf}
\begin{figure}
\epsscale{1.0}
\plotone{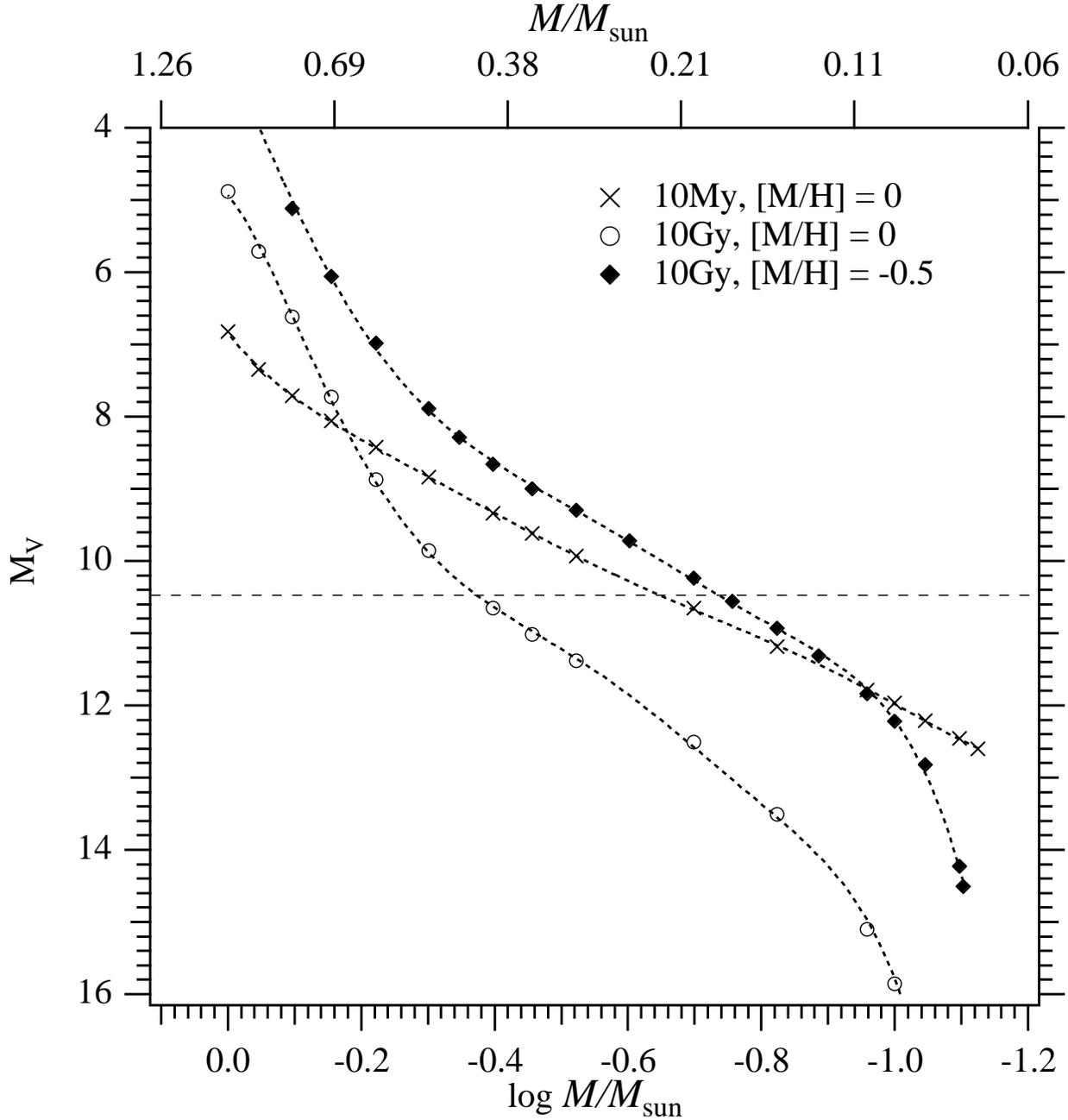}
\caption{M dwarf absolute magnitude as a function of mass from the stellar 
evolution models
of \cite{Brf98}. A wide range of masses, ages, and metallicities can result in
the derived dM absolute magnitude, $M_V  =  10.46$ (horizontal dashed line). Note that at a given mass, a low metallicity ([M/H]=-0.5) star is always brighter than a high metallicity ([M/H] = 0.0) star.} 
\label{fig-4}
\end{figure}

The dM absolute magnitude ($M_V = 10.46 \pm 0.12$ ) implies a spectral type M2V 
(\cite{Hen94}), consistent with \cite{Lie77}. The absolute magnitude of an M 
dwarf star depends not only on mass, but also on age (evolutionary stage) and
chemical composition. \cite{Brf98} have produced a grid of models, varying 
metallicity, [M/H] and helium abundance, Y.
We plot in Figure~\ref{fig-4} their Mass-Luminosity curves
 for dwarfs with ages 10My and 10Gy, with [M/H] = 0 and 10Gy with [M/H] = -0.5, all with solar helium abundance. 
The complete grid of Baraffe et al. models shows that M dwarfs in the mass range 
$0.175 \le {\cal M}_{dM} \le 0.43 {\cal M}_{\sun}$ with -0.5 $<$ [M/H] $<$ 0 have $M_V = 10.46$ at some 
time in their evolution from 10My to 10Gy.

\begin{figure}
\epsscale{1.0}
\plotone{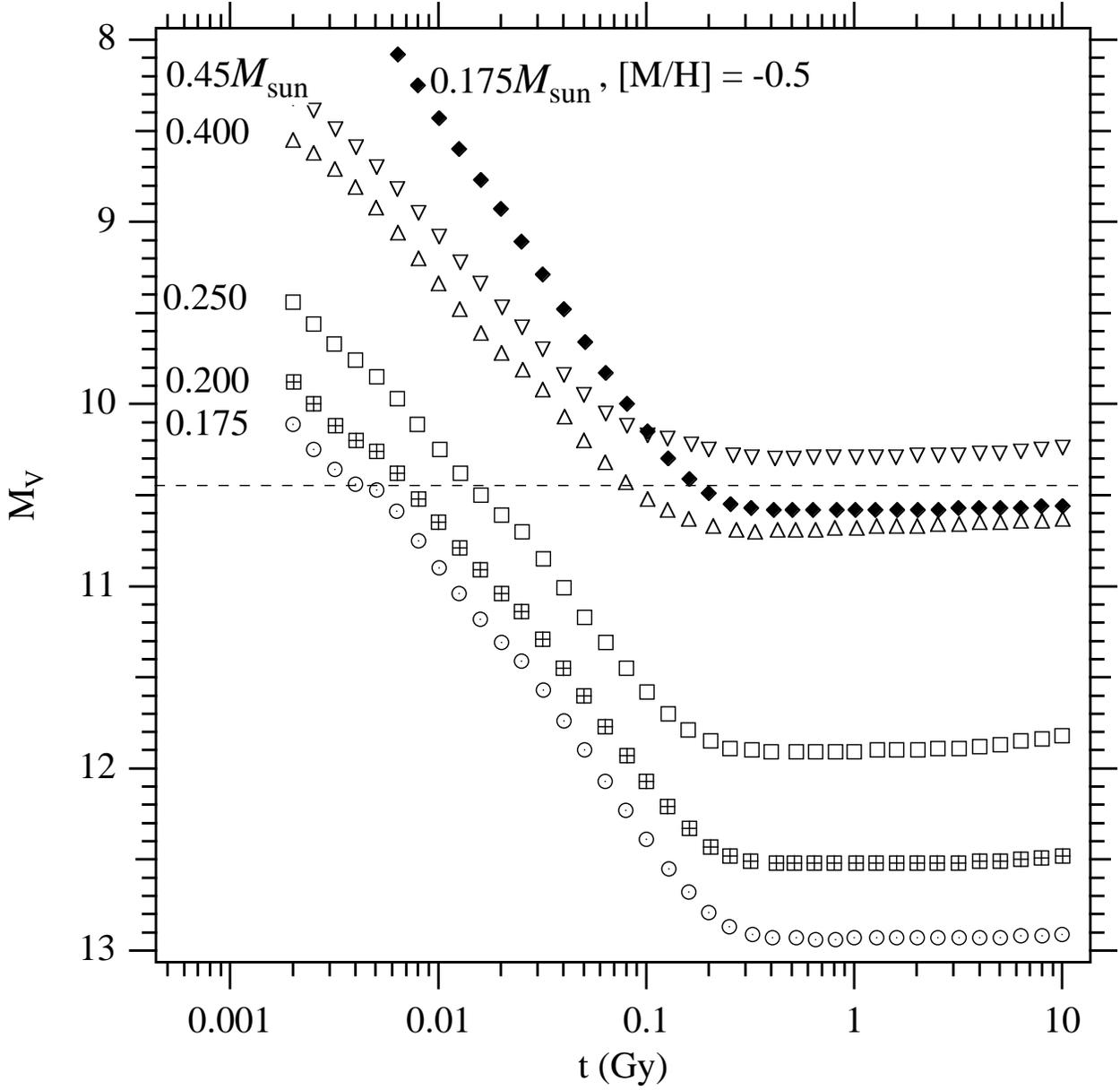}
\caption{The time variation of absolute magnitude for M dwarfs of various
masses taken from the stellar evolution models
of \cite{Brf98}. Empty symbols are for solar metallicity; filled for low metallicity ([M/H]=-0.5). Note that the higher mass stars remain near the computed M 
dwarf absolute magnitude, $M_V  =  10.46$ (dashed line), far longer than the low 
mass stars. A low metallicity star with ${\cal M}_{dM} = 0.175{\cal 
M}_{\sun}$ is brighter than a solar metallicity star with ${\cal M}_{dM} = 0.40{\cal 
M}_{\sun}$.} 
\label{fig-5}
\end{figure}

The dM mass now depends on metallicity and how quickly an M dwarf of a given mass decreases in 
brightness. Figure~\ref{fig-5} shows the dependence of brightness on mass, age, and metallicity.  These Baraffe et al. models indicate that solar metallicity stars with higher mass remain near
$M_V = 10.46$ far longer than low mass stars. However, kinematically, Feige 24 is more 
likely to be old and of lower than solar metallicity than young and of normal metallicity. First adopting the 10Gy model, [M/H] = 0, and calculated absolute 
magnitude, we estimate the dM star mass ${\cal M}_{dM} = 0.43 \pm 0.08 {\cal 
M}_{\sun}$, because that mass remains at $M_V = 10.46$ for a larger fraction of 
the total lifetime than any other. However, if we accept the kinematical suggestion of allegiance to a thick disk population, then [M/H] $<$ 0 is more likely. Assuming [M/H] = -0.5 results in a dM star mass ${\cal M}_{dM} = 0.185 \pm 0.08 {\cal M}_{\sun}$.  

Radial velocities from VT94 (dM from Kitt Peak, DA from IUE)
provide the velocity amplitude ratio, 
$K_{DA}/K_{dM} = 0.63 \pm 0.04 ={\cal M}_{dM}/{\cal M}_{DA}$. From the total possible dM mass range, 
$0.185 < {\cal M}_{dM} < 0.43 {\cal 
M}_{\sun}$, and the mass ratio we derive a DA mass range, 
$0.29 < {\cal M}_{DA} < 0.68 {\cal M}_{\sun}$. Applying the limit, ${\cal M}_{DA~Kepler} \ge 0.44{\cal M}_{\sun}$, we obtain  $0.44 < {\cal M}_{DA} < 0.68 {\cal M}_{\sun}$. Keplerian lower limits argue for a dM star mass $0.26 < {\cal M}_{dM} < 0.43 {\cal 
M}_{\sun}$, a range consistent with a metallicity slightly less than solar and an age in excess of 0.3 Gy (Figure~\ref{fig-5}).

\subsubsection{DA Mass from Atmospheric Parameters}

The dM star does not provide a particularly precise DA mass estimate. If one knows the surface gravity, $g$, and the radius, $R$, the mass can be obtained through
\beq
M = g R^2 / G
\eeq
where G is the gravitational constant. The quantity $log~g$ comes from analysis of the line profiles in spectra. Recent determinations include: \cite{Mar97}, $log~g = 7.53 \pm 0.09$; \cite{Kid91}, $log~g = 7.45 \pm 0.51$; \cite{Ven97}, $log~g = 7.2 \pm 0.07$; \cite{Fin97}, $log~g = 7.17 \pm 0.15$; and \cite{Bar98}, $log~g = 7.36 \pm 0.12$. The full range of the measures and Equation 3 yield the range of mass values ${0.21 \le \cal M}_{DA} \le 0.47  {\cal M}_{\sun}$. Applying the limit, ${\cal M}_{DA~Kepler} \ge 0.44{\cal M}_{\sun}$ eliminates nearly all of these mass determinations.  In this case our radius and the Kepler limit indicate that $log~g$ should be at the high end of these measures.

\subsection{The White Dwarf Composition}

\begin{figure}
\epsscale{1.0}
\plotone{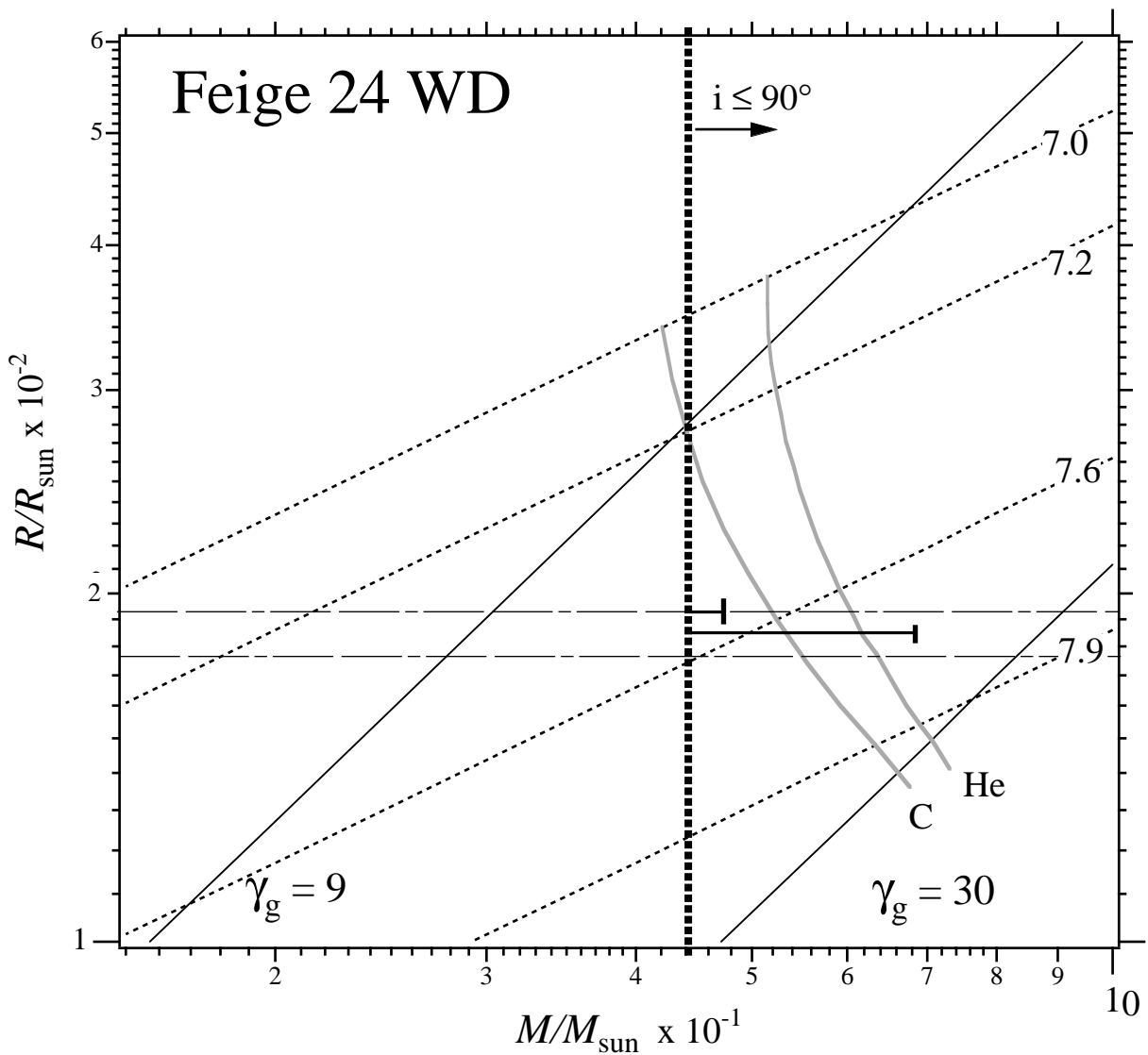}
\caption{Feige 24 mass and radius on a DA mass-radius map. The vertical bold dotted line shows the lowest possible ${\cal M}_{DA}$ from keplerian considerations. The radius error is represented by the top and bottom long-short dashed lines. The top thick horizontal bar shows the ${\cal M}_{DA}$ determined from atmospheric parameters. 
Only at the largest radius (lowest temperature) and largest $log~g$ do we obtain a DA mass in excess of the keplerian limit. The bottom thick horizontal bar at $R_{DA} = 0.0185 R_{\sun}$ indicates the ${\cal M}_{DA}$ range derived through the dM. We also plot 
several values of $log~g$ (dashed) and $\gamma_{grav}$ (thin solid). The grey wide solid lines are C and He DA models from \cite{Ven95}. A C core DA is somewhat more likely than an He core DA.} 
\label{fig-6}
\end{figure}

We next place Feige 24 on a white dwarf 
mass-radius diagram (Figure~\ref{fig-6}). We plot our two independently determined mass ranges against our adopted radius, $R_{DA} = 0.0185 \pm 0.0008 R_{\sun}$. We represent the radius error by the two horizontal long-short dashed lines. The top thick horizontal bar shows the ${\cal M}_{DA}$ determined from atmospheric parameters. Only the largest $log~g$ at the largest radius produces masses in excess of the Keplerian limit. The thick bar at $R_{DA} = 0.0185 R_{\sun}$ indicates the ${\cal M}_{DA}$ range derived through the dM mass estimates. For this determination the mass error bars indicate the range of ages and [M/H] discussed in section 3.2.2. For any dM older than 1-2 Gy the lower masses are associated with lower metallicity. The vertical bold dotted line shows the lowest possible ${\cal M}_{DA}$ that can produce the observed VT 94 radial velocity amplitudes for an edge-on orientation of this binary system. We also plot 
several values of $log~g$ (dashed) and $\gamma_{grav}$ (thin solid). 
The wide grey curves in Figure~\ref{fig-6} are C and He DA models from \cite{Ven95}. While uncertain, a carbon core DA seems more likely than a pure He core DA.

\section{Discussion}
While our estimated dM and DA masses differ little from VT94, our DA radius differs substantially. VT94 note the difference between their minimum 
radius, $R_{DA} = 0.028 R_{\sun}$, and that predicted by the \cite{Dah88} parallax. This 
discrepancy is exacerbated by the two new parallax determinations ({\it HST}
and {\it HIPPARCOS}), folded into our weighted average parallax. 

VT94 derive a DA gravitational redshift, $\gamma_{grav} = 8.7 \pm 2 ~km~s^{-1}$ 
from the measured mean velocities for the dM and DA.
Combined with our $R_{DA} = 0.0185 R_{\sun}$, this $\gamma_{grav}$ suggests 
a forbidden DA mass,  ${\cal M}_{DA}\sim 0.3 {\cal M}_{\sun}$. 
Reducing the mass of the DA component could reconcile the VT94 $log~g$ and  $\gamma_{grav}$ with our radius. 

We speculate that a third component in the Feige 24 system, a low-mass companion to the DA star, could preserve the total system mass and lower the DA mass. 
If all components are coplanar, the VT94 DA radial velocities apply strict limits to this reconciliation, because too high a mass for component C would show up as large residuals. We estimate from the scatter that a radial velocity amplitude of $\pm$10 \kms could 'hide' in the VT94 DA radial velocity measurements. Stellar dynamics applies yet another constraint.
Holman \& Wiegert (1999) parameterize the stability of tertiary companions as a function of stellar component A and B mass function, $\mu = {\cal M}_A/({\cal M}_A + {\cal M}_B)$, and AB binary orbit ellipticity, e. With e=0 and $\mu = 0.39$ we find (from their table 3) that component C must have an orbital semi-major axis less than 0.3 times that of AB.  

Insisting that ${\cal M}_{DA} = 0.30{\cal M}_{\sun}$ (this mass - with our radius -  would produce the upper limit VT94 $\gamma_{grav} = 10.7 $ \kms) requires ${\cal M}_{C} = 0.14{\cal M}_{\sun}$ (${\cal M}_A + {\cal M}_C = 0.44{\cal M}_{\sun})$. To hide the C component from the radial velocity technique requires a very low AC inclination, nearly face-on. However, non-coplanarity reduces the size of the stable AC semi-major axis even further (Weigert \& Holman, 1997; \cite{Pen83}). As an example suppose component C must have an orbital semi-major axis of 0.1 or less that of AB to insure stability. An AC period, P = 0.18$^d$ (4.3$^h$), and i = 6\arcdeg ~would produce a radial velocity signature of about $\pm 10$ \kms. Finally, the 
Mass-Luminosity Relation of Henry et al. (1999) would predict $M_V^C = 14.0$, hence, $V_C \sim 17.2$, likely undetectable in any of the spectra analyzed for radial velocities. Have we built a new CV, one that should evidence mass transfer and all the associated phenomena? A recent review of CVs (\cite{Beu99}) indicates that the putative component C (${\cal M}_{C} = 0.14{\cal M}_{\sun}$) would have to orbit much closer (P$\sim1.5^h$) to the DA primary before filling its Roche lobe and
producing the characteristic signature of a CV.

Finally we note that our radius differs little from that derived by VT94 from the only trigonometric parallax then available (\cite{Dah88}). The unresolved inconsistency between radii (derived from direct parallaxes) and surface gravities (derived from minimum mass and those radii) illuminates the need
for high angular resolution observations and direct mass determinations. 

The Feige 24 DA mass will rest on an age- and metallicity-dependent lower main sequence Mass-
Luminosity Relationship or still uncertain $log~g$ measurements until the component separations are
measured directly. Resolving the inconsistencies between the DA mass estimates (involving dM stellar models and 
uncertain temperatures, $log~g$, and bolometric corrections) requires
astrometry, both to further reduce the parallax uncertainty, and, more
importantly, to spatially resolve this system. Astrometrically derived 
orbital parameters will provide unambiguous and precise mass determinations for 
both components. They may also offer insight regarding past and future component interactions.

This system and dozens more like it are ideal targets for the Space 
Interferometry Mission (http://sim.jpl.nasa.gov). Feige 24, at a distance of 69 pc with P = 4.23$^d$, 
has a total component separation on order 700 microarcsec. The component orbits 
are much larger than the expected SIM measurement limits. Because shortward of 700 
nm 70-80\% of the system flux is  contributed by the DA (Thorstensen et al. 
1978), the wide SIM bandpass and spectral resolution should allow measurement of 
positions, magnitudes, and colors for both components, even with $\Delta V \sim 
2$. 

Once launched SIM will provide crucial astrometry for this and similar systems 
at ten times the distance (determined by target magnitude, not astrometric 
precision). SIM measurements of this system along with many other binaries will 
provide data with which to create an age- and metallicity-dependent Mass-Luminosity Relationship 
of exquisite accuracy.

\section{Conclusions}

1. The weighted average of three independent parallax mesurements yields a 
distance to the dM + DA binary Feige 24 with $\sigma_D / D$ = 2.8\%. D = $68.4^{+2.0}_{-1.9}$~pc.

2. We estimate the radius of the DA component using two methods. The first requires either a model-dependent bolometric correction, or one that derives from hot, normal stars. The second utilizes a model-dependent V-band average flux, $H_V$. The two results agree within their errors and yield a weighted average $R_{DA} = 0.0185 \pm 0.0008 R_{\sun}$, where the most significant contributions to the error are the uncertain $T_{eff}^{DA}$ and B.C.. This radius is larger than any of the WD discussed in \cite{Pro98}. 

3. FGS photometry provides quantitative evidence that
the orbital light curve is caused by heating of the dM component by the DA. That signature 
 is consistent with the assumed temperature and the radius we
have derived for the white dwarf 

4. The VT94 measured radial velocity amplitudes, amplitude ratios, and the assumption of Keplerian circular motion exclude ${\cal M}_{DA~Kepler} < 0.44 {\cal M}_{\sun}$ and ${\cal M}_{dM~Kepler} < 0.26 {\cal M}_{\sun}$.

5. We estimate the dM component mass, $0.26 < {\cal M}_{dM} < 0.43 {\cal 
M}_{\sun}$, from the \cite{Brf98} stellar evolution models,  a lower limit from Keplerian circular orbits, and the VT 94 radial velocities. The upper range is due to unknown age and metallicity, [M/H]. A DA mass range ($0.44 < {\cal M}_{DA} < 0.68 {\cal M}_{\sun}$) follows directly from the VT94 radial velocity amplitudes. 

  6. We determine ${\cal M}_{DA}$ from our $R_{DA}$ and a rather wide range of spectroscopically determined $log~g$ values. This approach yields ${0.44 \le \cal M}_{DA} \le 0.47  {\cal M}_{\sun}$, where again the lower limit is imposed by  ${\cal M}_{DA~Kepler} > 0.44 {\cal M}_{\sun}$.

7. We plot these DA component mass ranges on the ${\cal M}-R$ plane. With the assistance of the hard lower mass limit and C and He DA ${\cal M}-R$ models from \cite{Ven95},  we identify Feige 24 to have a carbon core. A pure He core DA seems less likely. 

8. Noting that our radius and the minimum possible ${\cal M}_{DA}$ are inconsistent with the VT94 $\gamma_{grav}$, we explore the possibility of a tertiary component. A component C, orbiting a common center of mass with the DA, having a period in the range  1.5 $<$P $<5^h$ with the orbit plane nearly face-on, could reduce the DA mass to ${\cal M}_{DA} = 0.30  {\cal M}_{\sun}$ and not produce any observational evidence.

8. SIM will be able to measure the orbits of each known component and provide directly measured dynamical masses for both. Orbit size and precise shape may provide information on the nature of past and future interactions between the two components. SIM would also detect a tertiary, if present.

\acknowledgments

This research has made use of NASA's Astrophysics Data
System Abstract Service and the SIMBAD Stellar Database inquiry and retrieval 
system. We gratefully 
acknowledged web access to the astrometry and photometry in USNO-A2.0, provided by the United States Naval Observatory, Flagstaff Station. Support for this work was provided by NASA through grant GTO 
NAG5-1603 from the Space Telescope 
Science Institute, which is operated
by the Association of Universities for Research in Astronomy, Inc., under
NASA contract NAS5-26555. We thank Don Winget and Sandi Catal\'{a}n for
discussions and draft paper reviews. Denise Taylor provided crucial scheduling
assistance at the Space Telescope Science Institute. Travis Metcalfe kindly provided
code for JD to HJD corrections. We thank an anonymous referee for suggestions that enhanced the clarity of our presentation, and for giving us the courage to speculate.
\clearpage


\clearpage

\begin{center}
\begin{deluxetable}{lcccccc}
\tablewidth{0in}
\tablecaption{Feige 24 Reference Frame: Astrometry  \label{tbl-2}}
\tablehead{  \colhead{ID}&\colhead{V}&   \colhead{$\xi$}&  
\colhead{$\eta$}&   \colhead{$\mu_X$}& \colhead{$\mu_Y$} &\colhead{$\pi$}
\\ & &   (arcsec)&  
(arcsec)& (arcsec y$^{-1}$)& (arcsec y$^{-1}$) &(arcsec)}

\startdata
ref-2*&11.59&0.0$\pm$0.0004&0.0$\pm$0.0004&0&0&0\nl
ref-3&13.38& -118.9943$\pm$0.0004& 50.4968$\pm$0.0004& 0.0157$\pm$0.0004 &-
0.0014$\pm$0.00050&-0.0006$\pm$0.0003\nl
ref-4&14.82&27.3168$\pm$0.0010&63.5709$\pm$0.0010&-
0&0&0\nl
ref-5&13.66&-144.4703$\pm$0.0004&86.9851$\pm$0.0005&0&0&0\nl
\enddata
\tablenotetext{*} {RA, Dec = 288.087967, 2.898281 (J2000)}
\end{deluxetable}
\end{center}

\begin{center}
\begin{deluxetable}{lccccccccc}
\tablewidth{0in}
\tablecaption{Feige 24 and its Reference Frame: Stellar Parameters  \label{tbl-3}}
\tablehead{  \colhead{ID}&
\colhead{ V \tablenotemark{a}}&   \colhead{\bv \tablenotemark{b}}&  \colhead{ 
SpT }&  
\colhead{$M_V$}&   \colhead{E(B-V) \tablenotemark{c}} &\colhead{$A_V$} 
&\colhead{m-M}&\colhead{D(pc)}&\colhead{$\pi_{abs}$}(mas)}
\startdata
ref-2&11.59&1.00&G9 III&0.75&0.03&0.093&10.84&960&0.6\\
ref-3&13.38&0.69&G3 V&4.8&0.03&0.093&8.58&410&1.9\\
ref-4&14.82&0.61&G0 V&4.4&0.03&0.093&10.42&935&0.8\\
ref-5&13.66&0.63&F9 III&1.2&0.03&0.093&12.46&2540&0.3\\
Feige 24&12.41\tablenotemark{d}&-0.20\tablenotemark{d}& & & & & & &\\
\enddata
\tablenotetext{a} {from FGS PMT measures calibrated as per \cite{Nel99}}
\tablenotetext{b} {from \bv = f(Sp.T.) + E(\bv)}
\tablenotetext{c} {from \cite{Bur82}}
\tablenotetext{d} {from \cite{Lan83}}

\end{deluxetable}
\end{center}

\begin{center}
\begin{deluxetable}{llrr}
\tablewidth{0in}
\tablecaption{{\it HST} Observations of Feige 24 and Astrometric Residuals  \label{tbl-4}}
\tablehead{  \colhead{Obs. Set}&
\colhead{MJD}&   \colhead{X residual}&  \colhead{Y residual}}
\scriptsize
\startdata
1&49930.92188&0.0002&0.0008\\
1&49930.9375&0.0000&-0.0012\\
1&49930.94531&-0.0003&-0.0004\\
2&49936.88672&0.0005&-0.0002\\
2&49936.90234&-0.0002&-0.0003\\
2&49936.91016&0.0006&0.0001\\
3&50102.09375&-0.0014&0.0008\\
3&50102.10938&0.0001&-0.0002\\
3&50102.11719&0.0004&-0.0013\\
4&50109.06641&-0.0008&0.0007\\
4&50109.07813&-0.0001&0.0002\\
4&50109.08594&0.0007&0.0001\\
5&50669.78125&-0.0001&0.0010\\
5&50669.79688&0.0009&0.0016\\
5&50669.80469&-0.0010&-0.0004\\
6&50678.92188&-0.0003&-0.0004\\
6&50678.9375&0.0003&0.0006\\
6&50678.94531&-0.0016&0.0010\\
7&50819.28125&0.0014&-0.0003\\
7&50819.28906&0.0005&-0.0008\\
7&50819.29688&0.0003&0.0001\\
8&50821.29688&0.0006&-0.0009\\
8&50821.30469&-0.0005&-0.0008\\
8&50821.3125&-0.0007&0.0006\\
\enddata
\end{deluxetable}
\end{center}

\begin{center}
\begin{deluxetable}{lll}
\tablecaption{Feige 24 Parallax, Proper Motion, and Radial Velocity \label{tbl-5}}
\tablewidth{0in}
\tablehead{\colhead{Parameter} &  \colhead{Value}}
\startdata
{\it HST} study duration  &2.4 y\\
number of observation sets    &   8 \\
Number of ref. stars &  4  \\
ref. stars $ <V> $ &  $13.4$  \\
ref. stars $ <B-V> $ &  $0.7$ \\
{\it HST} Relative Parallax & 13.8 $\pm$ 0.4  mas\\
corr to absolute & 0.9 $\pm$ 0.7   mas\\
{\it HST} Absolute Parallax & 14.7 $\pm$ 0.6   mas\\
{\it HIPPARCOS} Absolute Parallax &13.4 $\pm$3.6 mas \\
USNO Absolute Parallax& 13.5 $\pm$ 2.9  mas\\
{\it HST} Proper Motion ($\mu$)  &71.1 $\pm$ 0.6 mas y$^{-1}$ \\
 \indent in p.a. & 83\fdg6 \\
{\it HIPPARCOS} $\mu$  &85.8 $\pm$ 5 mas y$^{-1}$ \\
 \indent in p.a. & 84\fdg2 \\
USNO $\mu$  &78.4 $\pm$ 1.9 mas y$^{-1}$ \\
 \indent in p.a. & 88\fdg4 \\
Weighted Average Absolute Parallax & 14.6 $\pm$ 0.4   mas\\
m-M (LK bias corrected) & $4.17 \pm 0.11$\\
System Radial Velocity,  $\gamma$& + 62.0 $\pm$ 1.4 \kms &VT94\\
Galactocentric z velocity, W &  $-37 \pm 1.5$ \kms & $\gamma$ \& {\it HST} or {\it HIPPARCOS} $\mu$
\enddata
\end{deluxetable}
\end{center}

\begin{center}
\begin{deluxetable}{lll}
\tablecaption{Feige 24 Astrophysical Quantities \label{tbl-6}}
\tablewidth{0in}
\tablehead{\colhead{Parameter} &  \colhead{Value} &  \colhead{Source}}
\startdata
$ V_{tot} $    &   12.41$\pm$0.01 & \cite{Lan83}\\
\bv & -0.20 $\pm$ 0.01 & \cite{Lan83} \\
$V_{DA}$  & $12.56 \pm 0.05$ & \cite{Hol86}\\
$V_{dM}$  &   $ 14.63 \pm 0.05$ & $ V_{tot} $ \& $V_{DA}$\\
$A_V$ &$ 0 $ & reference frame $<B-V>$ \& Sp.T. (Table~\ref{tbl-3}) \\
m-M (LK bias corrected) & $4.17 \pm 0.11$ & this paper\\
dM $M_V $ & $ 10.46 \pm 0.12$& m-M\\
dM Sp. T. & M2V& dM $M_V$ \& \cite{Hen94}\\
${\cal M}_{dM} $ & $0.29 - 0.43 {\cal M}_{\sun}$& dM $M_V$,  
\cite{Brf98}, ${\cal M}_{DA~Kepler}$\\
DA $M_V$ & $8.39 \pm 0.12$ & m-M\\
DA B.C. & $-4.82 \pm 0.06$ & Flower, 1996 \\
$M_{bol}^{DA}$&  $3.57 \pm 0.13$& $= DA~M_V +B.C.$\\
$T_{eff}^{DA}$ & $56,370 \pm 1,000 K$ &  \cite{Bar98}\\
$R_{DA} $ & $ 0.0185 \pm 0.0008 R_{\sun}$& this paper\\
${\cal M}_{dM}/{\cal M}_{DA}$&$0.63 \pm 0.04$ & $K_{DA}/K_{dM}$  VT94 \\
${\cal M}_{DA}$ & $ 0.49^{+0.19}_{-0.05} {\cal M}_{\sun}$ &${\cal M}_{dM} $, 
$K_{DA}/K_{dM}$\& 
${\cal M}_{DA~Kepler}$ \\
${\cal M}_{DA}$ & $ 0.44 - 0.47 {\cal M}_{\sun}$ &$log~g $ \& 
${\cal M}_{DA~Kepler}$ \\
\enddata
\end{deluxetable}
\end{center}


%
%

\clearpage

\clearpage

\clearpage

\clearpage

\clearpage

\clearpage

\clearpage

\clearpage
\clearpage

\end{document}